\documentclass[aps,pra,twocolumn,groupedaddress,showpacs]{revtex4}

\usepackage{amsmath}

\begin{document}



\title{Dynamical symmetries of two-dimensional systems in relativistic quantum mechanics}

\author{Fu-Lin Zhang}
\email[Email:]{flzhang@mail.nankai.edu.cn} \affiliation{Theoretical
Physics Division, Chern Institute of Mathematics, Nankai University,
Tianjin 300071, People's Republic of China}

\author{Ci Song}
\affiliation{Theoretical Physics Division, Chern Institute of
Mathematics, Nankai University, Tianjin 300071, People's Republic of
China}

\author{Jing-Ling Chen}
\email[Email:]{chenjl@nankai.edu.cn}

\affiliation{Theoretical Physics Division, Chern Institute of
Mathematics, Nankai University, Tianjin 300071, People's Republic of
China}

\date{\today}

\begin{abstract}
The two-dimensional Dirac Hamiltonian with equal scalar and vector
potentials has been proved commuting with the deformed orbital
angular momentum $L$. When the potential takes the Coulomb form, the
system has an $SO(3)$ symmetry, and similarly the harmonic
oscillator potential possesses an $SU(2)$ symmetry. The generators
of the symmetric groups are derived for these two systems
separately. The corresponding energy spectra are yielded naturally
from the Casimir operators. Their non-relativistic limits are also
discussed.
\end{abstract}


\maketitle


Dynamical symmetries have been known for a long time in both
classical mechanics and quantum mechanics \cite{Greiner}. In
classical mechanics, both the orbits of the the Kepler problem and
the isotropic oscillator are closed \cite{Bertand}. This feature
suggests there are more constants of motion other than the orbital
angular momentum. They have been shown as the Rung-Lenz vector
\cite{Lenz,Pauli} in the Kepler problem and the second order tensors
\cite{Frakdin} in the isotropic oscillator. These conserved
quantities generate the $SO(4)$ and $SU(3)$ Lie groups respectively.
They are not geometrical but the symmetries in the phase space, and
are called dynamical symmetries. These symmetries lead to an
algebraic approach to determine the energy levels. Generally, the
$N-$dimensional (ND) hydrogen atom has the $SO(N+1)$ and the
isotropic oscillator has the $SU(N)$ symmetry.

In the relativistic quantum mechanics, the motion of
spin-$\frac{1}{2}$ particle satisfies the Dirac equation. A
dynamical symmetry does not exist in either the Dirac hydrogen atom
or the Dirac oscillator. Our question is: Do there exist two Dirac
systems which have the same dynamical symmetries as the
non-relativistic hydrogen atom and isotropic oscillator separately?
In both the non-relativistic limits of the two Dirac systems, there
exits a spin-orbit coupling in the Hamiltonians \cite{DiracOsc,QED}.
This suggests that the main reason of the breaking of dynamical
symmetries is the spin-orbit coupling. Therefore, in the Dirac
system which has the same dynamical symmetry as the non-relativistic
hydrogen atom or the isotropic oscillator, the spin and orbital
angular momentum should be conserved separately, and the potential
takes the Coulomb or harmonic oscillator form.

Neither the spin nor the orbital angular momentum commutes with the
Dirac Hamiltonian, even though in the free particle system. But, it
has been shown that, in the Dirac system with equal scalar and
vector potentials, the total angular momentum can be divided into
conserved orbital and spin parts as \cite{LS}
\begin{eqnarray}\label{L&S}
\ \overrightarrow{L}=\begin{bmatrix}
 \overrightarrow{l}&0\\
 0& U_p\overrightarrow{l}U_P
 \end{bmatrix},
\ \overrightarrow{S}=\begin{bmatrix}
 \overrightarrow{s}&0\\
 0& U_p\overrightarrow{s}U_P
 \end{bmatrix},
 \end{eqnarray}
where $\overrightarrow{l}=\overrightarrow{r} \times
\overrightarrow{p}$,
$\overrightarrow{s}=\frac{\overrightarrow{\sigma}}{2}$ are the usual
spin generators, $\overrightarrow{\sigma}$ are the Pauli matrices,
and
$U_p=U_p^{\dag}=\frac{\overrightarrow{\sigma}\cdot\overrightarrow{p}}{p}$
is the helicity unitary operator \cite{Hel}. The components of them
form the $SU(2)$ Lie algebra separately. Then, the Dirac
Hamiltonian, in the relativistic units, $\hbar=c=1$, takes the form
\begin{eqnarray}\label{H}
H=\overrightarrow{\alpha}\cdot\overrightarrow{p}+\beta M
+(1+\beta)\frac{V(r)}{2},
\end{eqnarray}
where $\overrightarrow{\alpha}$ and $\beta$ are the Dirac matrices,
and $M$ is the mass. Ginocchio \cite{U31,U32} has proved that the
Hamiltonian has the $U(3)$ symmetry in three-dimensional (3D) case,
when the potential $V(r)$ takes the harmonic oscillator form. And
the angular momentum given in (\ref{L&S}) are three of the eight
generators.

In this work we discuss the two-dimensional (2D) case. We choose
$\alpha_1=\sigma_1, \alpha_2=\sigma_2$ and $\beta=\sigma_3$, and
write the Hamiltonian in matrix form as
\begin{eqnarray}\label{H2}
\ H=\begin{bmatrix}
 M+V(r)&p_1-ip_2\\
 p_1+ip_2&-M
 \end{bmatrix}.
\end{eqnarray}
One can notice there is no the helicity unitary operator in the 2D
system. So we introduce a 2D version definition of the conserved
orbital angular momentum as
\begin{eqnarray}\label{L2}
\ L=\begin{bmatrix}
 \l&0\\
 0& B^{\dag}\frac{l}{p^2}B
 \end{bmatrix},
\end{eqnarray}
where $B=p_1-ip_2$, $B^{\dag}=p_1+ip_2$, and $l=x_1 p_2-x_2 p_1$ is
the usual orbital angular momentum. It is easy to prove the
commutation relation $[L,H]=0$.

The following question is: do there exist any other additional
conserved quantities when the potentials $V(r)$ are of some spacial
forms? We assume the constants of motion take the form as
\begin{eqnarray}\label{Q}
\ Q=\begin{bmatrix}
 Q_{11}&Q_{12}B\\
 B^{\dag}Q_{21}& B^{\dag}Q_{22}B
 \end{bmatrix}.
\end{eqnarray}
The commutation relation $[Q,H]=0$ requires the matrix elements must
satisfy the equations:
\begin{eqnarray}\label{Qeqn}
Q_{12} &=& Q_{21},\nonumber\\
\lbrack Q_{11},V(r)\rbrack +\lbrack Q_{12},p^2 \rbrack   &=& 0,\\
\lbrack Q_{12},V(r)\rbrack +\lbrack Q_{22},p^2 \rbrack   &=& 0,\nonumber\\
Q_{11} &=& Q_{12}(2M+V(r))+Q_{22}p^2.\nonumber
\end{eqnarray}
They are the same as the 3D case \cite{U32} (this suggests $V(r)$ of
the Eq. (1) in \cite{U32} should be $V(r)/2$). In the following
paragraphs we will give the solutions of (\ref{Qeqn}) with the
Coulomb and harmonic oscillator potentials in turn.

\textit{Hydrogen atom}. In non-relativistic hydrogen atom, the
constants of motion are the orbital angular momentum $l$ and two
components of the Rung-Lenz vector
\begin{eqnarray}\label{RL}
R_{i}=\frac{f_{i}}{2Mk}-\frac{x_{i}}{r}    ,\ \ \ i=1,2,
\end{eqnarray}
where $f_1=2p_2l-ip_1$, $f_2=-2p_1l-ip_2$, and $k$ is the parameter
in the Coulomb potential $V^h(r)=-\frac{k}{r}$. One can get the
following relations easily
\begin{eqnarray}\label{CR}
\lbrack f_{i},p^2 \rbrack = 0,\ \ \
\lbrack -\frac{x_i}{r},V^h(r) \rbrack = 0,\\
\frac{1}{2Mk}\lbrack f_{i},V^h(r) \rbrack + \frac{1}{2M} \lbrack
-\frac{x_i}{r},p^2 \rbrack = 0.\nonumber
\end{eqnarray}
Therefore, we can obtain the solutions of (\ref{Qeqn}) when the
potential $V(r)=V^h(r)=-\frac{k}{r}$
\begin{eqnarray}\label{QH}
\ Q^h_i=\begin{bmatrix}
 \ 2MR_i+\frac{kx_i}{r^2}&(-\frac{x_i}{r})B\\
 B^{\dag}(-\frac{x_i}{r})& B^{\dag}(\frac{1}{k}\frac{f_i}{p^2})B
 \end{bmatrix},\ \ \ i=1,2.
\end{eqnarray}
The commutation relations of the quantities are
\begin{eqnarray}\label{RH}
\lbrack L,Q^h_1 \rbrack=iQ^h_2,\ \ \lbrack L,Q^h_2 \rbrack=-iQ^h_1,\\
\lbrack Q^h_1,Q^h_2 \rbrack=-i\frac{4}{k^2}(H_h^2-M^2)L,\nonumber
\end{eqnarray}
and
\begin{eqnarray}\label{Q2}
(Q^h_1)^2+(Q^h_2)^2=\frac{H_h^2-M^2}{k^2}(4L^2+1)+(H_h+M)^2,
\end{eqnarray}
where
\begin{eqnarray}\label{Hh}
\ H_h=\begin{bmatrix}
 M-\frac{k}{r}&p_1-ip_2\\
 p_1+ip_2&-M
 \end{bmatrix},
\end{eqnarray}
is the Dirac Hamiltonian in Eq. (\ref{H2}) with $V(r)=V^h(r)$, The
deformed orbital angular momentum $L$ and $Q^h_i$ commute with
$H_h$, $[L,H_h]=0$, $[Q^h_i,H_h]=0$.

These results show that the 2D Dirac system with equal scalar and
vector potentials has the $SO(3)$ symmetry. The relations of the
generators can also be used to solve the energy levels of this
system. We define the normalized generators
\begin{eqnarray}\label{A}
A_1 &=& [-\frac{4}{k^2}(H_h^2-M^2)]^{-\frac{1}{2}}Q^h_1 ,\nonumber \\
A_2 &=& [-\frac{4}{k^2}(H_h^2-M^2)]^{-\frac{1}{2}}Q^h_2 ,\\
A_3 &=& L.\nonumber
\end{eqnarray}
Then,
\begin{eqnarray}\label{ACR}
[A_i,A_j]=i\epsilon_{ijk}A_k, (i,j,k=1,2,3).
\end{eqnarray}
The $SO(3)$ Casimir operator is given by
\begin{eqnarray}\label{CSO3}
C_{so3}=A^2_1+A^2_2+A^2_3=j(j+1),\ \ \ j=0,1,2,...
\end{eqnarray}
Inserting (\ref{Q2}) and (\ref{A}) into (\ref{CSO3}), one can get
the eigenvalues of the Hamiltonian $H_h$ as
\begin{eqnarray}\label{Eh}
E^{\pm}_h=\frac{\pm n^2-k^2}{n^2+k^2}M,\ \ \ n=2j+1=1,3,5,...
\end{eqnarray}
It takes the same form as the 3D case but different in the values of
$n$ \cite{3Dh}.

When $M \rightarrow \infty$, $H \rightarrow M$, the non-relativistic
limit of the energy levels is given by
\begin{eqnarray}\label{LmtEh}
E^{+}_h \rightarrow M-\frac{2k^2}{n^2}M,
\end{eqnarray}
the second term of which agrees with the non-relativistic results
\cite{NRH}. We can also get the non-relativistic limits of the
conserved quantities
\begin{eqnarray}\label{LmtQHh}
H_h-M & \rightarrow & \frac{H^2_h-M^2}{2M} \rightarrow
\begin{bmatrix}
 \frac{p^2}{2M}-\frac{k}{r}&0\\
 0&\frac{p^2}{2M}.
 \end{bmatrix},\\
\frac{Q^h_i}{2M} & \rightarrow & \begin{bmatrix}
 \ R_{i} &0\\
 0&\frac{1}{2Mk}f_{i}
 \end{bmatrix}.
\end{eqnarray}
The upper-left elements of the above matrices are nothing but the
non-relativistic hydrogen atom Hamiltonian and the Lung-Lenz vector,
and the lower-right ones are their limits when $k \rightarrow 0$.

\textit{Harmonic oscillator}. When the potential takes the harmonic
oscillator form, $V(r)=V^o(r)=\frac{1}{2}M\omega^2r^2$, the Dirac
Hamiltonian becomes
\begin{eqnarray}\label{Ho}
H_o=\begin{bmatrix}
 \ M+\frac{1}{2}M\omega^2r^2&p_1-ip_2\\
 p_1+ip_2&-M
\end{bmatrix}.
\end{eqnarray}
To get the conserved
quantities, we review some results in non-relativistic harmonic
oscillator. The constants of motion in non-relativistic case are the
orbital angular momentum $J_2=\frac{l}{2}$ and the second order
tensors
\begin{eqnarray}\label{J}
J_i=\frac{1}{2}(\frac{1}{M\omega}[pp]_i+M\omega[rr]_i),\ \ \ i=1,3,
\end{eqnarray}
where $[rr]_1=x_1x_2$, $[rr]_3=\frac{x^2_1-x^2_2}{2}$,
$[pp]_1=p_1p_2$ and $[pp]_3=\frac{p^2_1-p^2_2}{2}$. They satisfy
\begin{eqnarray}\label{CRo}
\lbrack \lbrack pp \rbrack_{i},p^2 \rbrack = 0,\ \ \
\lbrack \lbrack rr \rbrack_{i},V^o(r) \rbrack = 0,\\
\frac{1}{2M\omega}\lbrack \lbrack pp \rbrack_{i},V^o(r) \rbrack +
\frac{\omega}{2} \lbrack \lbrack rr \rbrack_{i},p^2 \rbrack =
0.\nonumber
\end{eqnarray}
Insert these relations into (\ref{Qeqn}), we can get the conserved
quantities in relativistic the harmonic oscillator potential as
\begin{eqnarray}\label{QO}
Q^o_i=\begin{bmatrix}
 \frac{4}{\omega}J_i+ \lbrack rr\rbrack_iV^o(r)& \lbrack rr
 \rbrack_i B\\
 B^{\dag} \lbrack rr \rbrack_i & \frac{2}{M\omega^2}  B^{\dag} \frac{\lbrack pp
 \rbrack_i}{p^2} B
\end{bmatrix},\  \ i=1,3.
\end{eqnarray}
The commutation relations of $L$ and $Q^o_i$ are
\begin{eqnarray}
\lbrack Q^o_1,L \rbrack=i2Q^o_3,\ \ \ \lbrack L, Q^o_3
\rbrack=i2Q^o_1,\\
\lbrack Q^o_3,Q^o_1 \rbrack=i2\frac{2}{M\omega^2}(H_o+M)L,\nonumber
\end{eqnarray}
which show the constants of motion construct the $SU(2)$ Lie
algebra.

The Casimir operator of the $SU(2)$ is
\begin{eqnarray}
C_{su2}&=&\frac{1}{4}\frac{M\omega^2}{2(H_o+M)}[(Q^o_1)^2+(Q^o_3)^2]+\frac{L^2}{4}\\
&=&s(s+1),\ \ \ s=0,\frac{1}{2},1,\frac{3}{2}... \nonumber
\end{eqnarray}
Then, the eigenvalues of the Hamiltonian (\ref{Ho}) are got as
$E^{-}=-M$ or $E^{+}_n$, which is the real root of the cubic
equation
\begin{eqnarray}\label{Eeqn}
\frac{(E^{+}+M)(E^{+}-M)^2}{2Mw^2}-(n+1)^2=0,
\end{eqnarray}
where $n=2s=0,1,2...$. It is coincided with the result in 3D case in
\cite{U31}.

When $M \rightarrow \infty$, $H \rightarrow M$, and the coefficient
of elasticity $M\omega^2$ keeps unchangeably, Eq. (\ref{Eeqn})
becomes the quadratic equation
\begin{eqnarray}\label{Eeqn2}
(E^{+}-M)^2-(n+1)^2\omega^2=0,
\end{eqnarray}
which leads the non-relativistic energy levels of the 2D harmonic
oscillator. At the same time, the operators give
\begin{eqnarray}\label{LmtQHo}
H_o-M & \rightarrow &
\begin{bmatrix}
 \frac{p^2}{2M}+\frac{M\omega^2}{2}r^2&0\\
 0&\frac{p^2}{2M}.
 \end{bmatrix},\\
\frac{\omega Q^o_i}{4} & \rightarrow & \begin{bmatrix}
 \ J_{i} &0\\
 0&\frac{1}{2M\omega}\lbrack pp \rbrack _{i}
 \end{bmatrix}.
\end{eqnarray}


In conclusion, we have shown that the 2D Dirac systems with equal
scalar and vector potentials have the $SO(3)$ symmetry with the
Coulomb potential, and the $SU(2)$ symmetry for the harmonic
oscillator potential. The nature of these symmetries are not
geometrical but dynamical. Their Hamiltonian can be expressed in
terms of the Casimir operators of the symmetry groups, which yield
the energy spectra straightway. In non-relativistic limit, the
eigenvalues lead the corresponding non-relativistic results. And the
upper-left elements of the operators coincide with their
non-relativistic counterpart accurately, while the lower-right ones
take their free particle limits.

Since Ginocchio has found the $U(3)$ symmetry in 3D case with the
harmonic oscillator potential, we can foretell our treatment can be
generalized to the ND Dirac system to find the $SO(N+1)$ symmetry of
the hydrogen atom and the $SU(N)$ symmetry of the harmonic
oscillator. Recently, Alhaidaria \textit{et. al.} \cite{D-KG} has
revealed the equivalence between the Dirac equation and the
Klein-Gordon equation with the equal scalar and vector potentials.
This suggests the dynamical symmetries should exist in the spin-$0$
system.

\begin{acknowledgments}
We thank S. W. Hu for his suggestions in grammar. This work is
supported in part by NSF of China (Grants No. 10575053 and No.
10605013) and Program for New Century Excellent Talents in
University. The Project-sponsored by SRF for ROCS, SEM.
\end{acknowledgments}

\bibliography{DynamicSymmetryDirac}

\end{document}